\documentclass[aps,prb,amsmath,amssymb,twocolumn,superscriptaddress,showpacs,floatfix]{revtex4-1}

\usepackage{graphicx}
\usepackage{dcolumn}
\usepackage{bm}

\usepackage{amsmath}
\usepackage{amssymb}

\begin{document}

\title{Electronic ferroelectricity and magneto-electric effect  in phase separated magnetic oxides}

\author{O.~G.~Udalov}
\affiliation{Department of Physics and Astronomy, California State University Northridge, Northridge, CA 91330, USA}
\affiliation{Institute for Physics of Microstructures RAS, Nizhny Novgorod, 603950, Russia}

\author{I.~S.~Beloborodov}
\affiliation{Department of Physics and Astronomy, California State University Northridge, Northridge, CA 91330, USA}

\date{\today}


\begin{abstract}
We consider phase separated states in magnetic oxides (MO) thin films. We show that these states have a non-zero electric polarization. Moreover, the polarization is intimately related to a spatial distribution of magnetization in the film. Polarized states with opposite polarization and opposite magnetic configuration are degenerate. An external electric field removes the degeneracy and allows to switch between the two states.    So, one can control electric polarization and magnetic configuration of the phase separated MO thin film with the external electric field. 
\end{abstract}

\maketitle

\section{Introduction}\label{Sec:Intro}

In the past, magnetic oxides (MO) attract a lot of interest due to the colossal magnteto-resistance effect (promising various applications) and intriguing many-body physics~[\onlinecite{RevModPhys.73.583,DAGOTTO20011,Ramirez_1997,RN34,PhysRevLett.77.175,HaghiriGosnet2003}]. Recently, MOs come in to focus of researchers as magneto-electric (ME) materials~[\onlinecite{Spaldin391,RN35,RN36,PhysRevB.73.134416,PhysRevLett.111.127601,PhysRevB.65.134402}] in whose magnetization is strongly coupled to electric polarization. The magneto-electric coupling promises numerous applications such as energy efficient magnetic memory cells~[\onlinecite{doi:10.1021/nl070465o}], logical devices~[\onlinecite{LI201819,RN37}], subwavelength antenna~[\onlinecite{doi:10.1002/adma.201203792}], micromechanical devices~[\onlinecite{doi:10.1021/nn5056332}] and biomedical systems~[\onlinecite{doi:10.1063/1.5007708}] inspiring interest of researchers. The coupling between magnetic and electrical degrees of freedom may appear due to a spin-orbit interaction~[\onlinecite{PhysRevLett.98.057601}] inducing the electric polarization in the MOs with a spiral magnetization. Strong ME effects also occur in the MOs due to a complicated interplay between an exchange interaction, the Jahn-Teller effects and the Coulomb repulsion~[\onlinecite{van_den_Brink_2008}]. Another mechanism is the strain-mediated coupling between magnetization and electric polarization in hybrid MO/ferroelectric (FE) systems~[\onlinecite{PhysRevB.87.094416,PhysRevB.75.054408,doi:10.1063/1.4726427,doi:10.1002/adma.200900278}]. Surface magnetization in the MOs can be controlled by an electric field via a charge accumulation~[\onlinecite{PhysRevLett.104.127202,doi:10.1002/adma.200904326,Vaz_2012}]. While various mechanisms of the coupling have been proposed so far, there is no ME effect efficient enough to be used in applications. Therefore, the search of novel types of ME coupling is still of great importance.

One of the key feature of MOs is the phase separation (PS)~[\onlinecite{RN34,DAGOTTO20039,Kagan1999,PhysRevB.78.155113,0305-4470-36-35-304,1063-7869-44-6-R01}] due to which the crystal is split into ferromagnetic (FM) regions where conduction electrons are localized and antiferromagnetic (AFM) regions with no electrons. The phase separation is intimately related to the colossal magneto-resistance effect~[\onlinecite{RN34,PhysRevLett.93.037203,DAGOTTO20011}]. In the present work, we address the question if the PS states contribute to the magneto-electric effect in the MOs. In a PS state the magnetic phase separation co-exists with the charge carrier localization. The FM regions are negatively charged while the AFM regions are positively charged. Therefore, one can control location of the FM and AFM regions with an electric field. Recently, local control of the phase separated state at a surface of a MO crystal with an electric field produced by a tip of a scanning microscope was discussed in Refs.~[\onlinecite{doi:10.1080/00150193.2017.1292823, Hernandez2015}]. 

\begin{figure}
	\includegraphics[width=0.9\columnwidth]{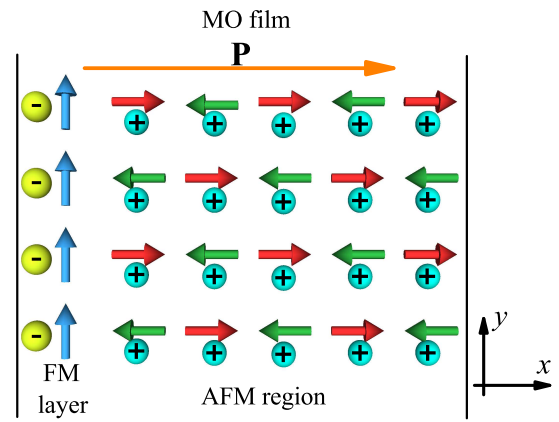}
	\caption{Phase separated magnetic oxide thin film with a single FM layer at the left surface and an AFM region in the rest of the film. Electrons are localized in the FM layer inducing a non-zero electric polarization $\mathbf P$. \label{Fig:Intro}}%
\end{figure}

In our work, we consider a thin film of MO and show that PS states are responsible for appearance of electric polarization (see Fig.~\ref{Fig:Intro}) and magneto-electric effect in such films. In particular, we demonstrate that the MO film is split into an FM layer and AFM one with an asymmetric charge distribution (non-zero electrical polarization). The polarization and locations of the FM and AFM layers can be switched with an electric field. 

The paper is organized as follows. Sec.~\ref{Sec:model} introduces our model of a MO thin film. The next section is devoted to discussion of different types of ground states in the MO film within a simplified analytical model. Sec.~\ref{Sec:Num} describes results of numerical simulations of the MO thin film. Appearance and electric field induced switching of polarization and magnetic configuration are discussed. Finally, we describe the limits of applicability of our model in Sec.~\ref{Sec:Disc}.

\section{The model}\label{Sec:model}

We consider a thin MO film with the thickness of $l$ atomic planes and the interatomic distance $\Delta$. The film is a perfect crystal with ideally flat surfaces. 
We use the most simple version of the double exchange model with a single orbital per site, ``classic'' magnetic moments and a cubic lattice to describe magnetic and electronic properties of the MO~[\onlinecite{Kagan1999,1063-7869-44-6-R01}].  The  Hamiltonian of the system reads
\begin{equation}\label{Eq:Ham}
\hat H=-\sum_{<i,j>} t_{ij}\hat a^+_i \hat a_j+\mathrm{C.C.}-J\sum_{<i,j>}\mathbf S_i\mathbf S_j+\hat U_{\mathrm{ion}}+\hat H_{\mathrm{C}},
\end{equation}
where $\mathbf S_i$ is the magnetic moment (normalized) of $i$-site, $J<0$  is the (AFM) intersite exchange coupling, $\hat a_i$ and $\hat a^+_i$ are the creation and annihilation operators for an electron at the site $i$, $t_{ij}$ is the transfer matrix element. This element depends on the mutual orientation of magnetic moments of sites $i$ and $j$, $t_{ij}=t\mathrm{cos}(\theta_{ij}/2)$, where $\theta_{ij}=\widehat{\mathbf S_i\mathbf S_j}$. Note that summation in both terms is performed over the nearest neighbours. Each site in the interior of the film has 6 neighbours. The surface sites have only 5 neighbours. The last two terms in Eq. (\ref{Eq:Ham}) describe the  Coulomb interaction of the electrons with each other, the electrons with ions, and the ions with each other. The ions potential is defined by the spatial distribution of dopands. We will consider the cases of homogeneous and inhomogeneous doping of the MO film. In the first case the density of ions $n_\mathrm i$ is homogeneous and equal to the electron density $n_0$. In the second case the ions density depends on the coordinate perpendicular to the film plane $n_\mathrm{i}(x)$ (keeping the whole system neutral). The Coulomb interaction can be written as
\begin{equation}\label{Eq:Coul}
\hat U_{\mathrm{ion}}+\hat H_{\mathrm{C}}=\frac{1}{2}\int\!\!\!\int d^3r_1 d^3r_2 \rho^\mathrm{tot}(\mathbf r_1)\rho^\mathrm{tot}(\mathbf r_2)/(|\mathbf r_1-\mathbf r_2|),
\end{equation}
where $\rho^\mathrm{tot}$ is the total charge density due to the ions and electrons.

\subsection{Numerical modelling procedure}

Let us introduce the site position vector $\mathbf r=(x,y,z)$ in which 
the coordinates $x$, $y$, and $z$ are measured in units of the lattice spacing $\Delta$. The $x$-coordinate is perpendicular to the film plane (see Fig.~\ref{Fig:Intro}). It varies from $1$ to $l$. 
The notation $t^{\pm x,y,z}_\mathbf r$ stands for the matrix element of the electron transfer from the site $\mathbf r=(x,y,z)$ to the  site neighbouring along the axis shown by the superscript $^x$, $^y$ or $^z$ and in the positive (superscript $^+$) or negative ($^-$) direction. The angles between neighbouring magnetic moments are denoted in a similar way $\vartheta^{\pm x,y,z}_\mathbf r$.

The system is uniform in the ($y$,$z$)-plane and $\vartheta_{\mathbf r}^{\pm x,y,z}$ and $t_{\mathbf r}^{\pm x,y,z}$ do not depend on these coordinates. At the film surface the transfer matrix elements are $t_{(1,y,z)}^{-x}=t_{(l,y,z)}^{+x}=0$.

A wave function of an electron is a plane wave in the $y$ and $z$ directions with a 2-dimensional wave vector $\mathbf k$ ($\Psi_\mathbf k=(\psi^{1}_\mathbf k,~\psi^{2}_\mathbf k,~...,\psi^{l}_\mathbf k)^Te^{i(k_y y+k_z z)}$). The wave function is a standing wave along the $x$-axis with the amplitude $\psi^{x}_\mathbf k$  at the site $x$. In this representation the Hamiltonian in Eq.~(\ref{Eq:Ham}) turns into the $l$ by $l$ matrix equation for the amplitudes $\psi^x_\mathbf k$ at a given wavevector $\mathbf k$, chosen magnetic configuration, and  distribution of electrical potential (see below). In our calculations we divide the $k$-space into 100$\times$100$\times$100 regions. 

The Coulomb interaction is taken into account via a self-consistent procedure. This procedure was used previously for modelling of MO/FE and MO/I interfaces and MO's superlattices~[\onlinecite{PhysRevB.84.155117, PhysRevB.73.041104,PhysRevB.80.125115,PhysRevB.78.024415}]. 
For a given magnetic moment configuration we fix the electron concentration $n_0$ (number of electrons $N$) during our calculations. We find a ground state step by step. At each step we calculate the electron density  distribution $\rho_\mathbf r$ and then the electric potential, $\Phi_\mathbf r$. Both the density and the potential are independent of $y$ and $z$. Therefore, we used the 1-dimensional discrete version of the Maxwell equation, $\Phi_{x+1}-2\Phi_x+\Phi_{x-1}=-4\pi\Delta^2(\rho_x-\rho_{\mathrm i})/\varepsilon$, where $\rho_x=e\sum_\mathbf k |\psi^{x}_\mathbf k|^2$ is the charge density of electrons, $\rho_\mathrm i=|e|n_\mathrm i$ is the charge density of ions, $\varepsilon$  is the MO dielectric constant, and $e$ is the electron charge. We consider the case of zero temperature in our simulations and only $N$ lowest levels are occupied. The electrical potential obtained at the current step is used at the next iteration for calculating of the electron wave functions.  Note that the interaction of the ions with the potential should be also calculated. These iterations are performed until the potential and the electron density converge to a certain solution. 

Magnetic energy is calculated straightforwardly from Eq.~(\ref{Eq:Ham}). 

Finally, we find the total energy $E^\mathrm{tot}$ for a given magnetic configuration. Varying the magnetic configuration we find the one having the lowest energy for given film parameters $l$, $J$, $U_0$, and $t$ ($U_0=4\pi e^2/(\varepsilon \Delta)$ is the characteristic Coulomb energy).

We also tune ionic potential $\hat U_\mathrm{ion}$ through the inhomogeneous doping $n_\mathrm i(x)$. See on this below.

\begin{figure}
	\includegraphics[width=1\columnwidth]{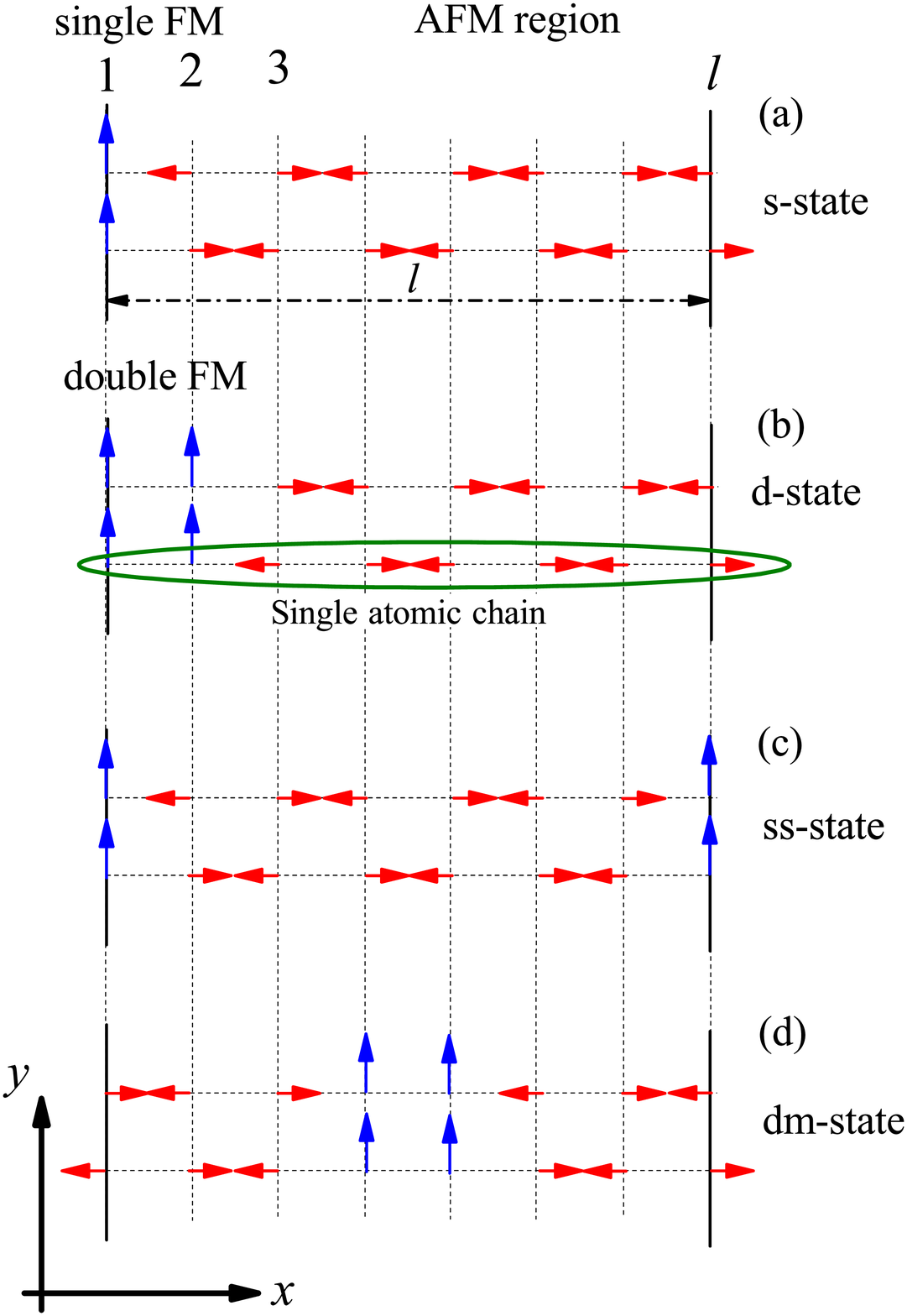}
	\caption{Four different configurations of magnetic moments in a MO thin film of the thickness $l$. The blue arrows show the magnetic moments  ordered ferromagnetically in the ($y$,$z$)-plane. The red arrows depict the magnetic moments directed along the $x$-axis. These moments are ordered antiferromagnetically in checker-board manner  (G-ordering). The numbers at the top enumerates atomic planes in the film. The two upper panels correspond to the asymmetric states with (a) one FM layer  or (b) two FM layers at the left surface. The two bottom panels show the states which are symmetric. They have single FM layer at the both surfaces (c) or a double FM layer in the middle of the film (d). The green ellipse captures a single atomic chain.  \label{Fig:SimpleStates}}%
\end{figure}

\section{Ground state of a magnetic oxide thin film}

\subsection{Uniform and phase separated states}

Generally, there are two types of ground states in the system. The first one is the state with uniform distribution of electrons across the film and constant angle between neighbouring magnetic moments, $\theta_{ij}=\theta_0$. The homogeneous state has symmetric charge distribution with zero polarization.

The second type is the PS state in which the system is split into the AFM region with $\theta_{ij}=\pi$ and the FM region with $\theta_{ij}=0$ (or the canted FM region with $\theta_{ij}=\theta_0<\pi$). At that, electrons are localized in the FM region while there are no electrons in the AFM one.  

The type of the ground state is defined by the competition between the Coulomb, kinetic, and magnetic energies. Separation into AFM and FM regions reduces the magnetic energy, but increases the kinetic and the Coulomb contributions. 

\subsection{Symmetric and asymmetric states}

The PS state with the inhomogeneous charge distribution may potentially have non-zero polarization. It was shown previously that phase separation in the form of 1-dimensional stripes can be stable in bulk MOs~[\onlinecite{Udalov2019}]. The question that we are addressing here is if such a stripe structure realizes in the MO film and if it takes symmetrical (see, for example, Fig.~\ref{Fig:SimpleStates}(c,d)) or asymmetrical (see Fig.~\ref{Fig:SimpleStates}(a,b)) position in the film. In the first case the charge distribution is an even function of coordinates and there is no electric polarization. In the second case the electric polarization appears. For example, if the FM stripe snuggles to one side of the film then this interface acquires a negative charge while the AFM region has a positive charge (see Fig.~\ref{Fig:Intro}). The electric polarization pointed from the FM stripe to the AFM region occurs. Such a polarized state is twice degenerate. Due to the symmetry of the film the FM layer can be at either of the interfaces. The energy of both states is the same but the polarization has the opposite sign. These degenerate states can be used as logical ``1'' and ``0''. An electric field should remove the degeneracy and allow to switch between ``1'' and ``0''. 

There are two different configurations of the PS film where polarization is zero. The first one is when the FM region is placed in the middle of the film (see Fig.~\ref{Fig:SimpleStates}(d)) forming symmetric charge distribution. The second one is when two FM regions appear at both edges of the film (see Fig.~\ref{Fig:SimpleStates}(c)). This configuration also gives an even charge distribution. In our modelling we compare the homogeneous, PS symmetric and PS asymmetric states and find the parameter region where the electric polarization may occur.

\subsection{A simplified consideration}

Here we use a simple analytical model to treat a few different magnetic configurations in the MO film. The first one has one atomic plane with the FM ordering at the edge of the film and the rest of the volume with the AFM ordering (see Fig.~\ref{Fig:SimpleStates}(a)). All electrons are in the FM plane and cannot transfer to the adjacent AFM layer. The single electron kinetic energy per a single atomic chain (see green ellipse in Fig.~\ref{Fig:SimpleStates}(b)) is given by 
\begin{equation}
E^\mathrm s_\mathrm k=-2t(\cos(k_y\Delta)+\cos(k_z\Delta)).
\end{equation}
We consider the limit of low electron concentration and use the effective mass approximation. The kinetic energy of all electrons per one atomic chain is given by
\begin{equation}
E_\mathrm k=-2t(2ln_0-\pi l^2n_0^2).
\end{equation}
The magnetic energy of a single atomic chain can be calculated as follows. The edge site at the left is in the FM plane. It has 4 neighbours in this plane giving the energy contribution $-4J$. The interaction with the magnetic moment in the second plane (AFM) is zero since the angle between the spins in the first and second atomic layers is $\pi/2$. The magnetic moment in the second layer interacts antiferromagnetically with 5 neighbours giving the energy contribution $5J$. The magnetic moment at the right interface also has  5 neighbours only. All internal magnetic moments have 6 neighbours and the magnetic energy of $6J$. Finally, the magnetic energy per one chain is given by
\begin{equation}
E_\mathrm m=\frac{1}{2}(-4J+5J+6J(l-3)+5J)=3J(l-2).
\end{equation}
One-half in front of expression is for avoiding of the double counting. 

We estimate the Coulomb energy assuming that the volume of the sample is split into two uniformly charged regions. A region of thickness $\Delta$ (single atomic layer at the left side of the film) is negatively charged (since all electrons are there) and the rest volume is positively charged (due to ions). The whole system is neutral. Then the Coulomb energy is given by
\begin{equation}
E_\mathrm C=\frac{U_0n_0^2 l(l-1)^2}{6}.
\end{equation}
The total energy of this state is
\begin{equation}\label{Eq:EnS}
E_\mathrm s^\mathrm{tot}=-2t(2ln_0-\pi l^2n_0^2)+3J(l-2)+\frac{lU_0n_0^2(l-1)^2}{6}.
\end{equation}
The subscript $_\mathrm s$ denotes the considered state with a single FM layer at an interface.

Now lets consider the configuration where two planes at opposite surfaces are FM ordered (see ss-state in Fig.~\ref{Fig:SimpleStates}(c)). The middle region is AFM. Using above approach we find
\begin{equation}\label{Eq:EnSS}
E_\mathrm{ss}^\mathrm{tot}=-2t\!\!\left(\!\!2ln_0-\frac{\pi l^2n_0^2}{2}\right)+J(3l-11)+\frac{lU_0n_0^2\left(\frac{l}{2}-1\right)^2}{3}.
\end{equation}

\begin{figure}
	\includegraphics[width=1\columnwidth]{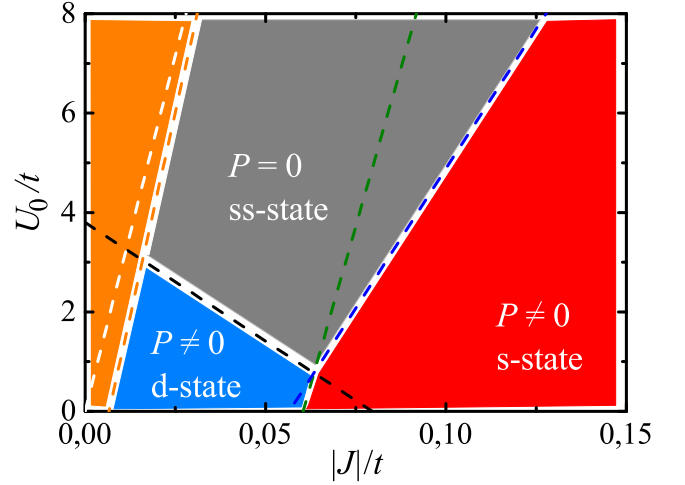}
	\caption{The phase diagram of a MO film within the simplified analytical model. The black dashed line shows Eq.~(\ref{Eq:dss}). The green line corresponds to Eq.~(\ref{Eq:sd}). The blue line is defined by Eq.~(\ref{Eq:sss}). The orange line shows the boundary beyond which (to the left) the states with triple FM layers should be taken into account. The white dashed line represents Eq.~(\ref{Eq:ddm}). \label{Fig:SimpleModel}}%
\end{figure}

The third configuration has two FM planes at an interface (see d-state in Fig.~\ref{Fig:SimpleStates}(b)). In this case the kinetic energy spectrum has two branches due to hopping of electrons between the two FM planes. Taking this into account we get for the total energy 
\begin{equation}\label{Eq:EnD}
E_\mathrm{d}^\mathrm{tot}\!=\!-2t\!\!\left(2ln_0-\frac{\pi l^2n_0^2}{2}+\frac{1}{8\pi}\!\!\right)+3J(l-4)+\frac{lU_0n_0^2(l-2)^2}{6}.
\end{equation}

The next state is the symmetric distribution with the double FM layer placed in the middle of the film (see md-state in Fig.~\ref{Fig:SimpleStates}(d)). We assume here that $l$ is even. The energy of this state is
\begin{equation}\label{Eq:EnDM}
E_\mathrm{dm}^\mathrm{tot}\!=\!-2t\!\left(\!2ln_0\!-\!\frac{\pi l^2n_0^2}{2}\!+\!\frac{1}{8\pi}\!\right)\!+3J(l-6)+\frac{lU_0n_0^2\left(\frac{l}{2}\!-\!1\right)^2}{3}.
\end{equation}

The configuration with the double FM layer is more favourable than the one with FM layers at the both interfaces  ($E_\mathrm{d}^\mathrm{tot}<E_\mathrm{ss}^\mathrm{tot}$) when
\begin{equation}\label{Eq:dss}
|J|<\frac{t}{4\pi}-\frac{lU_0n_0^2(l/2-1)^2}{3}.
\end{equation}
Single FM layer at a surface is more favourable than the double FM layer ($E_\mathrm s^\mathrm{tot}<E_\mathrm d^\mathrm{tot}$) when
\begin{equation}\label{Eq:sd}
|J|>\frac{1}{6}\left(\frac{t}{4\pi}+\pi t l^2 n_0^2+\frac{lU_0n_0^2(2l-3)}{6}\right).
\end{equation}
The state with a single FM layer is more favourable than the state with FM layers at the both interfaces ($E_\mathrm s^\mathrm{tot}<E_\mathrm{ss}^\mathrm{tot}$) when
\begin{equation}\label{Eq:sss}
|J|>\frac{1}{5}\left(\pi t l^2 n_0^2+\frac{lU_0n_0^2(l^2/2-1)}{6}\right).
\end{equation}
The symmetric state with a double FM layer in the middle of the film is more favourable than the state with double FM layer at an interface ($E_\mathrm{md}^\mathrm{tot}<E_\mathrm{d}^\mathrm{tot}$) when
\begin{equation}\label{Eq:ddm}
|J|<\frac{lU_0n_0^2(l/2-1)^2}{36},
\end{equation}
and more favourable than the state with a single FM layer at an interface when
\begin{equation}\label{Eq:}
|J|<\frac{1}{12}\left(t\pi n_0 l^2+\frac{t}{4\pi}-\frac{lU_0n_0^2(l^2/2-1)}{6}\right).
\end{equation}

Considering only above configurations we plot a digram of states in a MO thin film depending on $U_0$ and $J$ (see Fig.~\ref{Fig:SimpleModel}). Such a diagram presented in Fig.~\ref{Fig:SimpleModel} corresponds to $n_0=0.05$ and $l=6$. For each of these states there is an area in the phase diagram  where the state is the most favourable. The dashed lines in  Fig.~\ref{Fig:SimpleModel} show Eqs.~(\ref{Eq:dss}), (\ref{Eq:sd}), (\ref{Eq:sss}) and (\ref{Eq:ddm}). 

The symmetric state with two single FM layers at the both interfaces of the film (ss-state) is the most favourable in the gray area. Electric polarization in this case is zero. 
\begin{figure*}[t]
\begin{center}
\includegraphics[width=2.05\columnwidth]{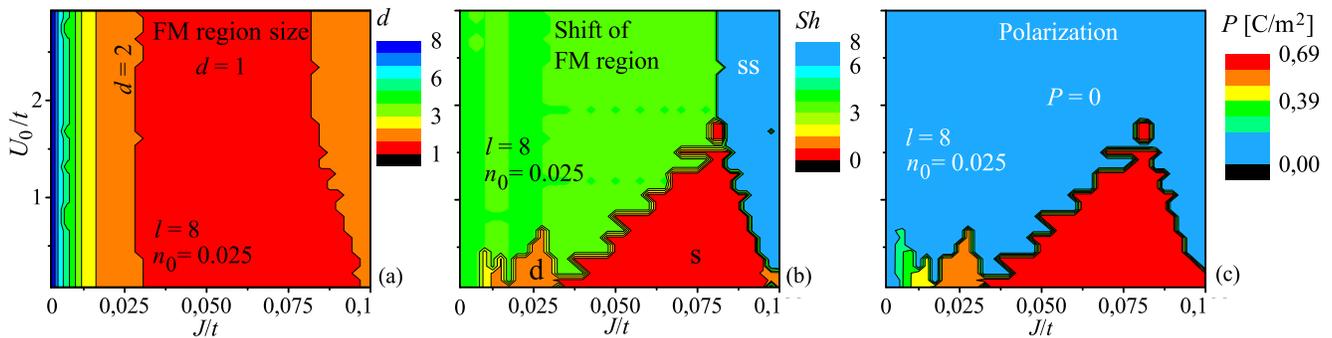}
\caption{Phase diagram of a MO film with the thickness $l=8$ atomic layers and electron concentration $n_0=0.025$. (a) Dependence of the FM region size $d$ corresponding to the state with the lowest energy on the characteristic exchange interaction $J$ and the Coulomb energy $U_0$. (b) Position  $Sh$ of the FM region in the film as a function of $J$ and $U_0$. The symbols ``s'', ``ss'', and ``d'' stand for the states with a single FM layer at a surface, FM layers at both surfaces, and a double FM layer at a surface, correspondingly.  (c) Electric polarization  of the film $P$.  \label{Fig:NumPD1}}
\end{center}
\end{figure*}

The blue area shows the parameters region in which the double FM layer at an interface (d-state) is the state with the lowest energy. Since the FM region is negatively charged and the AFM region is positively charged, the film acquires an electric polarization. Since the FM region can be at either of the surfaces this state is twice degenerate. Therefore, the polarization can be pointed either in positive or negative direction along the $x$-axis. 

The red colour shows the parameters region where the state with a single FM layer snuggled at an interface has the lowest energy. This configuration also has a non-zero electric polarization $P$ and is also twice degenerate. 

The smaller the exchange constant $J$ the bigger should be the FM region. So, at a certain small $J$ the double FM layer should enlarge to a triple FM layer. Within the same model (as we considered above) the transition to the state with the triple FM layer happens to the left of  the dashed orange line in Fig.~\ref{Fig:SimpleModel} (orange region). Beyond this line additional states should be involved into consideration.

The white dashed line corresponds to Eq.~(\ref{Eq:ddm}). To the left of this line the state with the double FM layer in the middle of the film is the most favourable. Since this line is inside the orange region this state is not realized in the system.

Note that the simplified model also fails for large $J$. In this limit the FM state turns into the canted FM state. This is not taken into account in the model.

\section{Numerical simulation}\label{Sec:Num}

The simplified model considered above shows that there is a region of parameters where the film acquires an electric polarization. However, several simplifications were used in the model which can be avoid in numerical simulations. In contrast to the simplified model the boundary between the FM and AFM regions is treated correctly, allowing electron to penetrate into the surface of AFM layer. The Coulomb interaction is treated self-consistently taking into account the influence of the electron-electron interaction onto the wave functions. Another important point is that the magnetic moments in the FM region can be in the canted FM state. In contrast to the simplified model a much wider range of states is studied when searching for the ground state. The magnetic configuration in the film is characterized by three parameters that are optimized during the simulations. The first one is the width $d$ of the FM region. $d$ varies from 1 to $l$, where $d=1$ gives a single FM layer and $d=l$ corresponds to uniform magnetic film with uniformly distributed electron density. The second parameter is the position of the FM region $Sh$ varying from $1$ to $l$. The positions $Sh=d/2$ and $Sh=l-1-d/2$ mean that the FM region is placed at the left or right interface, correspondingly. The shift of the FM layer $Sh=l$ gives magnetic configuration in which a half of the FM region is at the left interface and another half is at the right interface. At that, the AFM region is in the middle of the film. The last parameter we optimize is the angle between neighbouring magnetic moments in the FM region. We find the energy minimum varying these three parameters for given $n_0$, $U_0/t$, $J/t$ and $l$.

Figure~\ref{Fig:NumPD1} shows the phase diagram of the MO film of thickness $l=8$ with electron concentration $n_0=0.025$.

The parameter region correlates to real
materials constants: hopping matrix element $t = 0.1-0.4$ eV, the intersite Coulomb interaction $U_0 = 1-10$ eV, and $J = 0.01 - 0.1t$~[\onlinecite{Izyumov:2001,PhysRevLett.92.157203}].

The FM region size $d$, the shift $Sh$, and the electric polarization $P$ are shown. The electric polarization is calculated as the dipole moment perpendicular to the film plane per unit area
\begin{equation}\label{Eq:Polarizatio}
P=\frac{1}{\Delta^2}\sum_x \rho_x (x-x_\mathrm{c}),
\end{equation}
where $x_\mathrm c=l/2$ is the middle point of the film. To estimate the polarization we take $\Delta=0.5$ nm.

\begin{figure}
	\includegraphics[width=1\columnwidth]{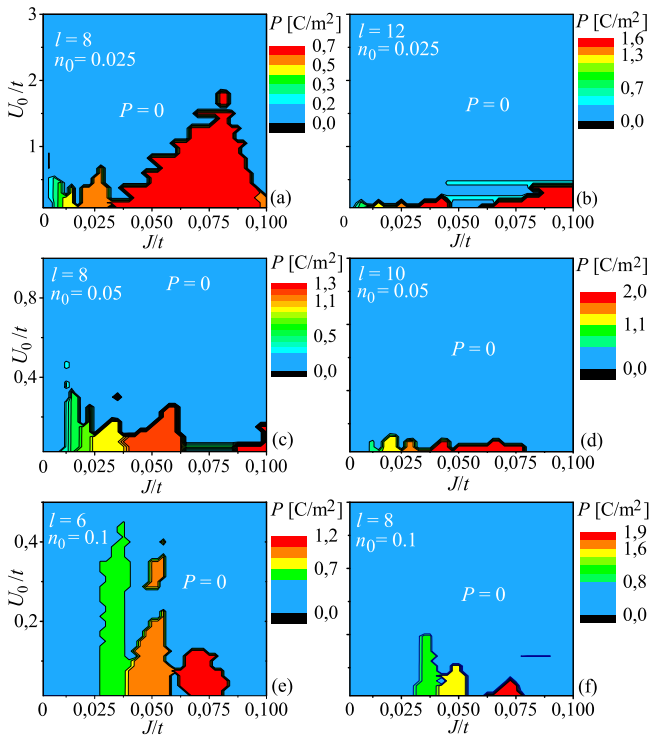}
	\caption{Polarization $P$ as a function of $U_0$ and $J$ for the MO thin films with different thickness $l$ and electron concentration $n_0$ (shown in panels). \label{Fig:Polarization}}%
\end{figure}

Left panel in Fig.~\ref{Fig:NumPD1} shows the FM region size for the state with the lowest energy as a function of $U_0$ and $J$ normalized by $t$. One can see that the size of the FM region gradually decays with increasing of the exchange interaction $J$ as expected. At small $J$ the size of the FM region is equal to the film thickness meaning that the uniform state is the most favourable and there is no phase separation. When $J$ is high enough the FM region shrinks to 1 layer. The central panel shows the shift $Sh$ of the FM region corresponding to the minimum energy of the system. The shift $Sh=4$ (green area) corresponds to symmetric position of the FM region in the middle of the film. While in the simplified analytical model this configuration is not realized, here we see that such a state is the most favourable in practically whole parameters region. The blue area in the upper right corner of the phase diagram corresponds to the state with the FM region  symmetrically split to both surfaces (ss-state, $Sh=l$). The red region at the bottom shows the states with an asymmetric position of the FM region. These states should have an electric polarization. This is shown in the right panel demonstrating $P(U_0,J)$. One can see that the states with finite polarization correspond to asymmetric states in the central panel. Comparing the left and right panels we can see that non-zero polarization occurs when the FM region size is 1 or 2. This is in agreement with our simplified analytical model. Moreover, the shape of the parameter regions in which $P\ne 0$ resembles what is shown in Fig.~\ref{Fig:SimpleModel}. The magnitude of the polarization is quite high. It reaches 0.6 C/m$^2$ which is just twice lower than the polarization in strong FEs such as BaTiO$_3$.

The dependence of polarization on the thickness $l$ and electron concentration $n_0$ is shown in Fig.~\ref{Fig:Polarization}. The region of parameters where the polarization exists decreases with increasing of the film thickness $l$ and electron concentration $n_0$. This happens because growing of $l$ and $n_0$ increases the Coulomb energy of the polarized state.

The magnitude of polarization grows with the film thickness since the polarized state is mostly a single FM layer at a surface of the film. Similarly, polarization grows with the electron density $n_0$.

\begin{figure}
	\includegraphics[width=1\columnwidth]{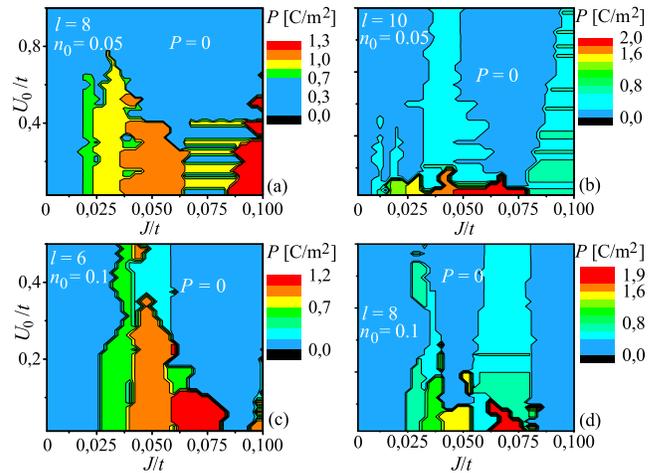}
	\caption{Polarization $P$ as a function of $U_0$ and $J$ for the MO thin films with different thickness $l$ and electron concentration $n_0$ in the case of inhomogeneous doping. The width of the undoped layer is $d_\mathrm{undoped}=2$ atomic planes. There are no donors in this region ($\Delta n_1=n_0$).  \label{Fig:PolInhDop}}%
\end{figure}

\subsection{Increasing the polarization using an inhomogeneous doping}

The polarization is zero when a spatially symmetric configuration is more favourable than the asymmetric one. There are two possible symmetric configurations with the FM layer in the middle of the film (see the green area in Fig.~\ref{Fig:NumPD1}(b)) and with two separate FM layers at the both surfaces of the film (see the blue area in Fig.~\ref{Fig:NumPD1}(b)). 

One can ``struggle against'' the fist type of the symmetric configurations using inhomogeneous doping. The electron concentration is defined by the amount of donors. If the layers close to the film surface are doped but the layers in the middle of the film are not doped, then an inhomogeneous ion potential occurs in the system and a potential hill for electrons appears in the middle of the film. This increases the energy of the symmetric states with the FM region placed in the centre of the system making polarized states more favourable. We model this situation by introducing spatially dependent background ionic charge into the model. In the middle of the film we make a region of the width $d_\mathrm{udoped}$ in which the positive charge density is lower than the ionic charge density outside of this region. Note, that we keep the whole system neutral. In the central region  the ionic charge density decreases by $\Delta n_1$. Outside this layer the ionic charge density increases by $\Delta n_2=\Delta n_1d_\mathrm{undoped}/(l-d_\mathrm{undoped})$.

Figure~\ref{Fig:PolInhDop} shows the polarization for several $l$, $n_0$, $d_\mathrm{undoped}$, and $\Delta n_1$. The inhomogeneous doping indeed increases the region of parameters where the non-zero polarization occurs (compare 2 bottom panels in Fig.~\ref{Fig:Polarization} and Fig.~\ref{Fig:PolInhDop}). 
 
\begin{figure}
	\includegraphics[width=1\columnwidth]{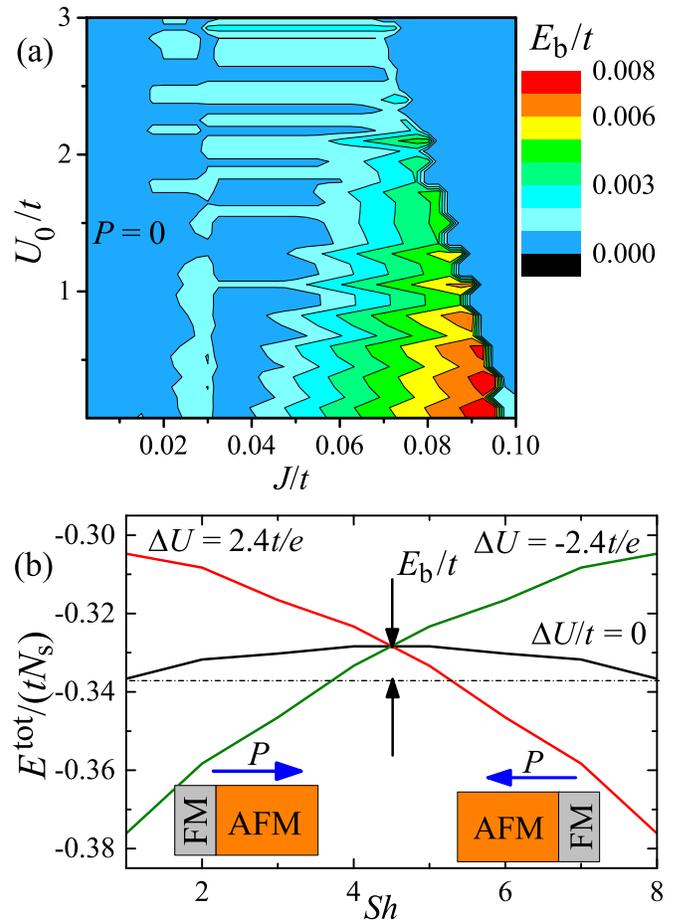}
	\caption{(a) Dependence of the switching barrier $E_\mathrm b$ between two states with opposite polarization on $U_0$ and $J$ for the film with thickness $l=8$ and electron density $n_0=0.025$. There is an inhomogeneous doping in the film with $d_\mathrm{undoped}=2$ and $\Delta n_1=-0.05$. (b) Dependence of the film energy $E^\mathrm{tot}$ (per one site, $N_\mathrm s$ is the number of sites in the system) on the position of the FM region $Sh$. The film parameters $l=8$, $n_0=0.025$, $U_0/t=1$, $J/t=0.75$. The black line shows the case of zero external electric field. The ground state of the system is the asymmetric s-state with a single FM layer at a surface ($Sh=1$ or $Sh=8$).  The red and green lines show the case of positive and negative applied external electric field. $\Delta U=El$ is the voltage applied across the film. The barrier height $E_\mathrm b$ is shown. Insets show polarized states corresponding to the configurations with a single FM layer at an interface. \label{Fig:BarAndME}}%
\end{figure}

\subsection{Polarization switching and magneto-electric effect}

In the previous section we showed that at certain parameters the MO thin film becomes electrically polarized. Due to the system symmetry the electrically polarized states are twice degenerate. Polarization can be pointed in positive or negative direction (FM layer can be either at left or right side). There is a question if the system  spontaneously switches between the two states. To address this question we find the energy of the system as a function of FM region position. In the polarized state the FM region is located at the 
interface. We assume that switching of polarization goes through motion of the FM  region across the film to the other surface. The system energy  as a function of the FM region position $E^\mathrm{tot}(Sh)$ has a peak when the FM region is in the middle of the film. A typical dependence $E^\mathrm{tot}(Sh)$ is shown in Fig.~\ref{Fig:BarAndME}(b) with the black line. To switch polarization the system should overcome the barrier $E_\mathrm b=E(l/2)-E(Sh_\mathrm g)$, where $Sh_\mathrm g$ is the FM region position corresponding to the ground state and the first term is the energy of the state with the FM region in the middle of the film.  We study how this barrier behaves as a function of $U_0/t$ and $J/t$. Fig.~\ref{Fig:BarAndME}(a) shows $E_\mathrm b$ for the case of $n_0=0.025$, $l=8$ and inhomogeneous doping in the region $d_\mathrm{udoped}=2$ and $\Delta n_1=0.05$. The barrier can be as high as $E_\mathrm b=0.01t$. Typical transfer matrix element $t\sim 1$ eV giving the barrier height of order of $0.01$ eV$=150$ K). So, the polarization can be stable below this temperature. The barrier can be increased or decreased by the proper doping of the middle layer. Generally, the barrier decreases with increasing $l$, $n_0$ and $U_0$, similarly to polarization.

The question of the barrier height is closely related to the electric field induced switching of the polarization in the MO film. Without an electric field the states with opposite electric polarization are degenerate. When we apply the electric field the degeneracy is removed. We calculate the energy of the system $E^\mathrm{tot}$ as a function of the FM region position  $Sh$ when the electric field $E$ is applied (the voltage drop across the film can be found as $\Delta U=El$). Figure~\ref{Fig:BarAndME}(b) shows these dependencies. One can see that in the strong enough electric field the barrier is removed and the system can be transferred to a state with electric polarization pointed to the electric field direction. Changing the polarity of the electric field one can switch the polarization in the phase separated MO.

The applied electric field switches not only the polarization but also the position of the FM region. So, in the phase separated MO film the magnetic state can be tuned with an electric field which is the consequence of the strong internal coupling between electric and magnetic degrees of freedom in these materials.

\section{Discussion}\label{Sec:Disc}

In this section we discuss limitations of our model of MO. 

1) The Jahn-Teller effects~[\onlinecite{Kugel1982}] are not taken into account. 
Generally, the JT effects should localize electrons~[\onlinecite{RN55}] and increase 
the stability of the polarized states discussed in our work.

2) Usually, in MOs several orbitals are vacant at a single site. As shown in Ref.~[\onlinecite{PhysRevLett.92.157203}], the strong JT interaction leads to large splitting of $e_g$ orbitals leading to formation of narrow band of polarons (localized) and conductive band. Our model can be considered as the limiting case of the model in Ref.~[\onlinecite{PhysRevLett.92.157203}] in which we neglect the narrow band of localized polarons.

Generally, depending on the doping regime different systems describing by a single or double orbital models can be obtained~[\onlinecite{PhysRevLett.82.1016}]. In the two-orbital models the complicated interplay of the orbital ordering, charge ordering and exchange interaction could induce a FE polarization in bulk MO crystals (not speaking about various additional magnetic phases). Introducing many-orbital model would be an interesting development of our model but requires a separate study. 

3) We do not take into account a disorder in our MO films. Such a disorder can appear due to randomness of  donors positions or due to localized JT polarons. This randomness can destroy the regular one-dimensional structures studied in this work. This issue requires an additional investigation.

4) We also neglect correlation effects. These effects are important for the case of high electron density. The on-site Coulomb interaction (entering the Hubbard correlation term) can be estimated as $U_\mathrm H\approx 5 - 10$ eV~[\onlinecite{Khomskii1973}]. The average energy due to the on-site repulsion is $W_\mathrm H=n_0^2U_\mathrm H$. Comparing this contribution to the characteristic kinetic energy $\sim  tn_0$, we 
estimate the concentration where correlation effects are negligible comparing to hopping, $n_0<6t/U_\mathrm H$. For $t\approx 0.5$ eV we get $n_0<0.1$.  Therefore, we focus here on the limit of small electron concentration. Note, however, that while the average concentration of electrons in our model is small, the concentration in the FM region can be much higher. For example, if all the electrons in 10 monolayers are placed in a single surface layer the electron concentration in this layer increases 10 times. So, even for small average concentration the correlation effects can be important in the case of PS states. This question requires further investigation.

\section{Conclusion}

We studied properties of MO thin films and showed that in a certain range of parameters the ground state of the system is the state with asymmetric charge distribution and therefore with non-zero electric polarization. In particular, the MO film is split into negatively charged FM region and positively charged AFM region. These regions are placed at the opposite sides of the film leading to the appearance of the electrical dipole moment. Such a ground state is twice degenerate. The two degenerate states have opposite electric polarization and opposite position of the FM region. Using an external electric field one can switch between these two degenerate states. So, one can control the electric polarization and the magnetic state of the phase separated MO thin film with the external electric field.

\section{Acknowledgements}

This research was supported by NSF under Cooperative Agreement Award EEC-1160504. 
O.~U. was supported by the Foundation for the
Advancement of Theoretical Physics and Mathematics “BASIS” (grant 18-1-3-32-1) and Russian Foundation for Basic Researches (Grant  18-32-20036).

\appendix

\bibliography{CDW}

\end{document}